# Studies on charge production from $Cs_2Te$ photocathodes in the PITZ L-band normal conducting radio frequency photo injector

C. Hernandez-Garcia[1], M. Kraslinikov, G. Asova[2], M. Bakr[3], P. Boonpornprasert, J. Good, M. Gross, H. Huck, I. Isaev, D. Kalantaryan, M. Khojoyan[4], G. Kourkafas[5], O. Lishilin, D. Malyutin[6], D. Melkumyan, A. Oppelt, M. Otevrel[7], G. Pathak, Y. Renier, T. Rublack, F. Stephan, G. Vashchenko[8], and Q. Zhao[9]

*Deutsches Elektronen Synchrotron DESY, Platanenallee 6, 15738 Zeuthen, Germany*

**Abstract**

This paper discusses the behavior of electron bunch charge produced in an L-band normal conducting radio frequency cavity (RF gun) from $Cs_2Te$ photocathodes illuminated with ps-long UV laser pulses when the laser transverse distribution consists of a flat-top core with Gaussian-like decaying halo. The produced charge shows a linear dependence at low laser pulse energies as expected in the quantum efficiency limited emission regime, while its dependence on laser pulse energy is observed to be much weaker for higher values, due to space charge limited emission. However, direct plug-in of experimental parameters into the space charge tracking code ASTRA yields lower output charge in the space charge limited regime compared to measured values. The rate of increase of the produced charge at high laser pulse energies close to the space charge limited emission regime seems to be proportional to the amount of halo present in the radial laser profile since the charge from the core has saturated already. By utilizing core + halo particle distributions based on measured radial laser profiles, ASTRA simulations and semi-analytical emission models reproduce the behavior of the measured charge for a wide range of RF gun and laser operational parameters within the measurement uncertainties.

---

[1] Corresponding author on leave from JLab, Newport News, VA 23606, USA, chgarcia@jlab.org
[2] on leave from BAS INRNE, 1784 Sofia, Bulgaria
[3] on leave from Assiut University, 71515 Assiut, Egypt
[4] now at Synchrotron SOLEIL, Paris, France
[5] now at Helmholtz-Zentrum Berlin, Berlin, Germany
[6] now at Helmholtz-Zentrum Berlin, Berlin, Germany
[7] Now at the Central European Institute of Technology, Brno, Czech Republic
[8] now at DESY, Hamburg, Germany
[9] on leave from IMP/CAS, 730000 Lanzhou, China

# Introduction

The Photo Injector Test facility at DESY, Zeuthen site (PITZ) [1], is dedicated to the development and optimization of high-brightness electron sources for free-electron lasers (FELs), such as FLASH [2], and the European XFEL in Hamburg that require electron bunches with extremely small transverse emittance [3]. Production of such electron bunches imposes stringent operational settings on the photoinjector that often lead to electron emission near the space charge (SC) limit.

The PITZ photoinjector consists of a normal conducting L-band radio frequency (RF) 1.6-cell cavity gun with a main-bucking focusing solenoid pair and a $Cs_2Te$ photocathode, a normal conducting RF booster cavity, a transport line with electron beam diagnostics, and a photocathode UV laser system with associated beam transport and diagnostics. By means of a temporal pulse shaper, the laser system can be tuned to generate from short (~2 ps FWHM) Gaussian pulses to long (~20 ps FWHM) flattop pulses, allowing the RF gun to produce bunch charges up to a few nC at maximum momentum of 7 MeV/c. The photoinjector optimization in 2008–2009 for bunch charges of 1, 0.5, 0.25, and 0.1 nC resulted in measured emittance values which met the requirements of the European XFEL [4]. With further improvements of the PITZ photoinjector in 2010-2012 even smaller emittance values were achieved, albeit a rather large discrepancy was observed between measured and simulated projected transverse emittance as a function of laser rms size at the cathode [5]. The optimum laser spot size at the cathode corresponding to the minimum projected transverse emittance measurements is smaller compared to that predicted by ASTRA beam dynamics simulations [6], despite direct plug-in of experimental parameters [5]. The discrepancy in optimum laser spot size is almost negligible for bunch charges below 0.1 nC, but becomes larger with increasing bunch charge.

Although the transverse phase space is one of the critical electron beam quality benchmarks demonstrated at PITZ for the European XFEL, this is a higher order beam dynamics effect compared to the bunch charge production. Charge measurements as a function of laser pulse energy consistently show that the accelerated charge continues to increase asymptotically in the space charge limited emission regime, while ASTRA simulations show charge saturation despite direct plug in of experimental laser and RF gun parameters. This work focuses on studying the sources for this discrepancy. Observations of the transverse laser distribution illuminating the photocathode indicate the presence of halo extending beyond the flattop core, rather than the ideal flattop radial distribution intended to be generated with the laser beam transport system. Although the simulations show that the charge saturates in the core using a uniform input particle distribution, radial halo may contribute to the measured additional extracted charge, but this is not observed in the simulations as long the ideal flattop laser transverse distribution is utilized.

# Experimental setup and procedures

*Temporal and transverse profile of the laser pulses*

The photocathode laser system provides UV pulses with a wavelength of 257 nm and a maximum energy of ~10 μJ per micro-pulse by means of an Yb:YAG regenerative amplifier and a two-stage Yb:YAG booster amplifier in combination with frequency conversion crystals [1, 2 and 7]. The system is capable of generating pulse trains containing up to 800 micro-pulses separated by 1 μs at 10 Hz repetition rate. The temporal shaping of the micro-pulses takes place in the laser room, before the laser beam line (LBL) that transports the beam to the photocathode RF gun in the accelerator tunnel. The flexibility of the laser system allows production of Gaussian pulses in a variety of lengths, from ~2 ps FWHM with no manipulation, up to ~11 ps FWHM with a Lyot filter in the regenerative amplifier. In addition, the Gaussian pulses can be transformed into a temporal flattop profile with rise and fall times as short as ~ 2 ps and pulse lengths between 17 and 24 ps FWHM by means of a longitudinal pulse shaper based on 13 birefringent crystals [7]. The temporal profile of the UV output pulses was characterized with an optical sampling system (OSS) based on an optical cross-correlation technique with resolution better than 1 ps [7].

The transverse shaping of the UV laser pulses is performed at the LBL towards the photo cathode [1]. The laser spot at the conversion crystals is imaged onto a Beam Shaping Aperture (BSA). The BSA is imaged through a vacuum window and reflected off a vacuum mirror at nearly normal incidence onto the RF gun photocathode for producing a homogeneous radial distribution. The diameter of the BSA can be finely adjusted with a remotely controlled iris diaphragm, while its position relative to the laser beam path can be set also remotely via stepper motors. Remote controllable mirrors allow centering the laser spot on the photocathode, thereby aligning the laser spot center of mass with the electrical axis of the gun cavity. The transverse distribution of the cathode laser is monitored with a UV sensitive CCD camera placed at a location optically equivalent to the real cathode position. The laser pulse energy delivered to the photocathode can be adjusted remotely via a rotatable half-wave plate followed by a birefringent crystal used as a polarizer. A pick-off mirror (3.6% reflectivity) directs a fraction of the laser beam to an energy meter. The actual laser pulse energy on the photocathode is calculated from this measurement taking into account ~91% transmission of the vacuum window, and ~85% reflectivity of the vacuum mirror.

*Charge measurements as a function of laser pulse energy*

The experiments consisted in measuring the produced charge as a function of laser pulse energy. The charge was measured with a Faraday cup located ~0.8 m downstream of the photocathode. The cathode accelerating RF field is given by

$$E_{cath} = E_0 \sin(\varphi) \quad (1)$$

where $\varphi = \Phi_0 - \Psi$, $E_0$ is the peak longitudinal component (accelerating) of the electric RF field, $\Phi_0$ is the zero-crossing phase and $\Psi$ is the gun set-point (SP) phase. Practically, the phase offset $\Phi_0$ is a parameter determined within the Low Level RF system and thus is dependent on its setup, e.g. on the peak RF power in the gun as well as resonance conditions of the cavity. The nominal operation phase of the RF gun is set for the Maximum Mean Momentum Gain (MMMG) of electron beam. The zero-crossing phase was estimated by setting the bunch charge to about 10 pC (significantly below space charge limited emission for the nominal operation phases) and measuring the charge as a function of gun phase until the charge saturated due to the quantum efficiency limited emission [1]. The gun phase that corresponds to the point where the extracted charge is ½ of the saturated charge value is the zero-crossing phase, as shown by the phase scan in figure 1. The uncertainty in the estimation depends on the effect of the image charge on the phase scan and on phase jitter (laser pulse arrival time w.r.t. the gun RF launch phase). Table 1 indicates the parameter space for each experimental setup.

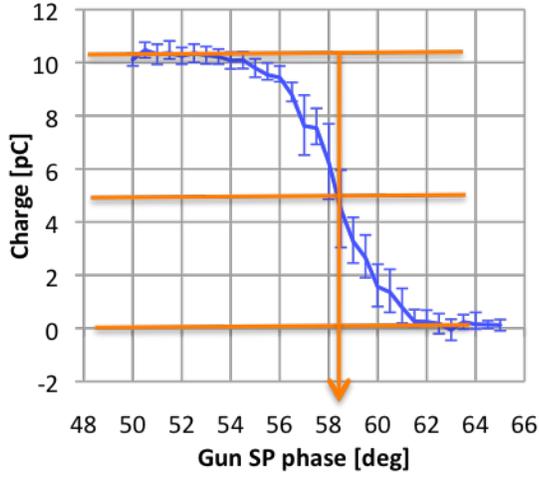

**Figure 1.** Typical phase scan for determining the zero-crossing phase $\Phi_0$, indicated by the arrow, corresponding to the phase for which the charge is ½ of the maximum as indicated by the arrow.

**Table 1**: Laser and RF gun parameters utilized for each experimental setup. MMMG is the RF gun phase corresponding to the Maximum Mean Momentum Gain. $E_{cathode}$ is calculated using equation 1. The laser pulse length was measured with the OSS for setups 1-4, while an extrapolation of earlier measurements with YLF Lyot filters (6 mm ~ 4 ps FWHM, and 16 mm ~ 7 ps FWHM) would suggest a FWHM pulse length of ~3.5ps FWHM [7 and 8] for setups 5-7. For setups 8-10 there was no Lyot filter in the laser regenerative amplifier; therefore the Gaussian laser pulse is estimated to be about 2 ps FHWM.

| Setup | BSA diameter (mm) | Laser temporal profile | Laser pulse length FWHM (ps) | Gun RF power (MW) | Gun RF Phase (deg) | $E_{cathode}$ at moment of emission (MV/m) |
|---|---|---|---|---|---|---|
| 1 | 1.2 | Gaussian | 2.7 | 4.000 | MMMG | 29 |
| 2 | 1.2 | Gaussian | 2.7 | 7.750 | MMMG | 45 |
| 3 | 1.2 | Flattop | 17.0 | 4.000 | MMMG | 29 |
| 4 | 1.2 | Flattop | 17.0 | 7.750 | MMMG | 45 |
| 5 | 0.8 | Gaussian | 3.5 | 1.500 | 90 | 29 |
| 6 | 0.8 | Gaussian | 3.5 | 3.375 | 90 | 43.5 |
| 7 | 0.8 | Gaussian | 3.5 | 6.000 | 90 | 58 |
| 8 | 0.8 | Gaussian | 2.0 | 6.000 | 90 | 58 |
| 9 | 0.8 | Gaussian | 2.0 | 6.000 | 49 | 43.5 |
| 10 | 0.8 | Gaussian | 2.0 | 6.000 | 30 | 29 |

## Results and discussion

Figure 2 shows the temporal profiles corresponding to setups 1-4 measured with the optical sampling system.

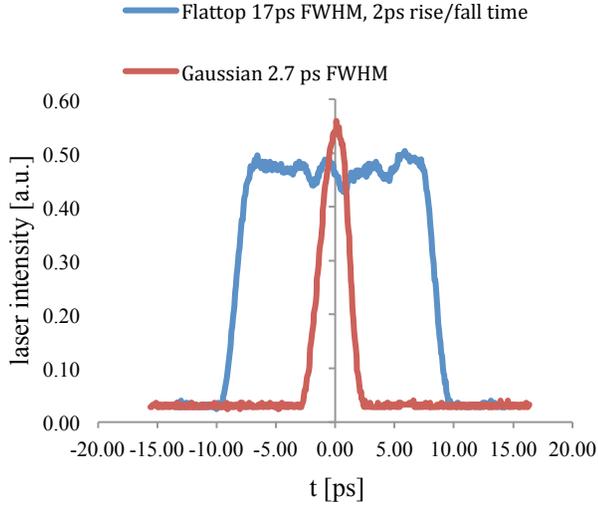

**Figure 2.** Laser temporal profiles measured with the OSS for setups 1-4.

The laser spot size on the cathode for setups 1-4 was established using the BSA set to 1.2 mm for both types of temporal pulses. The laser diode pumps were readjusted during tuning of the laser temporal profile. As a result a larger laser energy range was applied for the case of the flattop profile. This can explain a difference in the overall intensity between both distributions shown in figure 3. Note that the measured rms transverse size is ~0.312 mm compared to 0.300 mm expected from an ideal flattop core radial profile with 1.2 mm diameter.

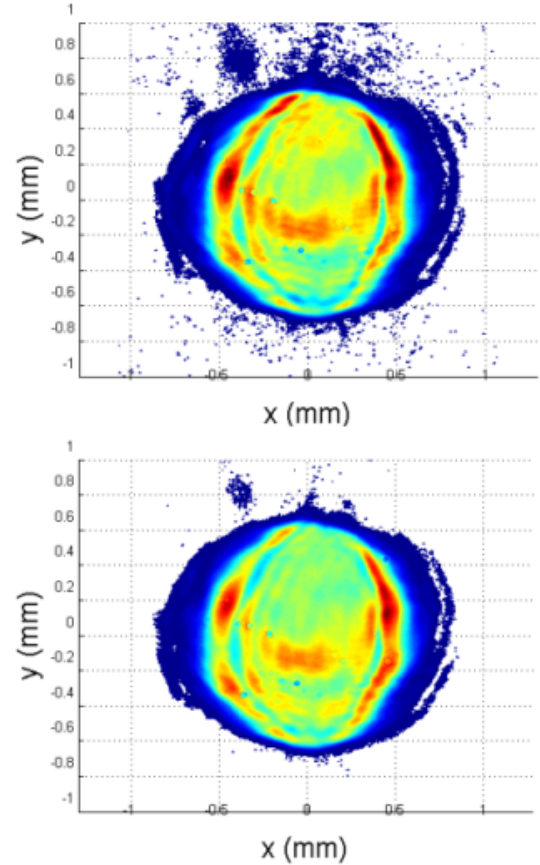

**Figure 3.** Laser transverse distribution images for BSA=1.2mm captured with a UV-sensitive CCD camera placed at a location optically equivalent to the real cathode position for the flattop temporal profile (top) with measured $\sigma_{xy}$=0.313 mm and for the Gaussian temporal profile (bottom) with measured $\sigma_{xy}$=0.312 mm.

The charge measurements vs. laser pulse energy for setups 1-4 shown in figure 4 were taken with RF gun power settings of 4.0 MW ($E_0$=45.9 MV/m) and 7.75 MW ($E_0$=62.7 MV/m) at the Maximum Mean Momentum Gain (MMMG) gun phase. Measured momentum distributions at the MMMG phase for these power settings have center of mass at 5.32 MeV/c and 7.09 MeV/c correspondently. Beam dynamics simulations yield cathode accelerating fields at the moment of emission of 29 and 45 MV/m, respectively. The MMMG phase was determined by measuring the beam momentum as a function of gun phase at the low energy dispersive arm [5]. For setups 1 and 2 the estimated rms phase jitter was ~2.5°, whereas for setups 3 and 4 the rms jitter was ~1.8°.

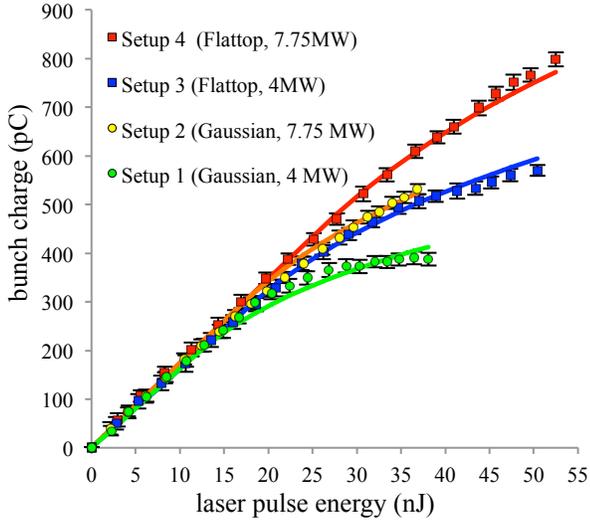

**Figure 4.** Measured charge vs. laser pulse energy for setups 1-4. The solid lines show the results of the semi-analytical model (equations 2-6) applied to the corresponding measurements.

Charge measurements vs. laser pulse energy show the strongest saturation for setup 1 in the space charge (SC) limited emission regime while those for setup 4 show the most linear behavior. In the quantum efficiency (QE) limited emission regime (lower laser pulse energy) the behavior for setups 1-4 is linear as expected. The laser transverse distributions in figure 3 suggest the presence of halo (dark blue) extending beyond the core (green), which exhibits intensity variations (red hot spots) in a ring-like fashion. Such laser radial profile can be represented to a first approximation by a homogeneous core distribution (without considering the intensity fluctuations within the center part) and a decaying halo outside the core with the following equation

$$F(r) = \frac{E_l}{\pi R_c^2 + 2\pi\xi\sigma_r^2} \begin{cases} 1, r \le R_c \\ \xi \cdot e^{\left(\frac{R_c^2 - r^2}{2\sigma_r^2}\right)}, r > R_c \end{cases}, \quad (2)$$

where $E_l = 2\pi \int_0^\infty F(r) r \, dr$ is the laser pulse energy, $R_c$ is the radius of the core, $\xi$ is the relative intensity of the Gaussian halo with respect to the intensity of the core, and $\sigma_r$ is the rms size of the Gaussian halo profile.

The surface space charge density assuming a radially homogeneous core with Gaussian-like decaying halo can be described by

$$\sigma_Q(r) \approx \frac{2 \cdot QE \cdot E_l}{\pi R_c^2 + 2\pi\xi\sigma_r^2} \begin{cases} 1, \text{ if } r \le R_c \\ \xi \cdot e^{\left(\frac{R_c^2 - r^2}{2\sigma_r^2}\right)}, \text{ if } r > R_c \end{cases} \quad (3)$$

where the approximate factor 2 comes from the calculation of the QE given in % $\left(QE = \frac{hc}{e\lambda} \cdot \frac{Q}{E_l}\right)$ for $\lambda$=257 nm, and considering that $E_l$ is given in nJ and the bunch charge $Q$ in pC. A simple model to estimate the effect of halo contributing to extracted charge beyond saturation in the core can be applied to our laser radial profile [9]. By taking into account the space charge density limit $\sigma_{scl} \propto E_0 \sin\varphi_0$ [5] and denoting $Q_{max} = \sigma_{scl} \cdot (\pi R_c^2 + 2\pi\xi\sigma_r^2)$ as the limiting charge value, assuming a space charge limitation the produced charge can be calculated as:

$$Q = Q_{core} + Q_{halo}, \quad (4)$$

where the core charge is given by:

$$Q_{core} = \frac{1}{1+\xi\eta} \begin{cases} Q_{exp}, \text{ if } Q_{exp} \le Q_{max} \\ Q_{max}, \text{ if } Q_{exp} > Q_{max} \end{cases}. \quad (5)$$

The halo charge can be calculated as:

$$Q_{halo} = \frac{\eta}{1+\xi\eta} \begin{cases} \xi Q_{exp}, & \text{if } \xi Q_{exp} \le Q_{max} \\ Q_{max} \cdot \left(1 + \ln\frac{\xi Q_{exp}}{Q_{max}}\right), & \text{if } \xi Q_{exp} > Q_{max} \end{cases} \quad (6)$$

where $Q_{exp}$ given in pC is the theoretically expected charge which would be emitted without presence of space charge forces (QE

limited emission regime), and $\eta = \frac{2\sigma_r^2}{R_c^2}$ for the halo-core rms area ratio. Indeed, if both core and halo are not saturated and ($Q_{exp} \leq Q_{max}$ and $\xi \cdot Q_{exp} \leq Q_{max}$), the total charge is $Q = Q_{core} + Q_{halo} = Q_{exp}$. The model can be applied to the simultaneous fit of the measured four curves (Fig. 4) by using six parameters: $Q_{max}$(Flattop), $Q_{max}$(Gaussian), $\xi$, $\eta$, $QE$(7.75MW) and $QE$(4 MW). For the space charge limit the following formula is used:

$$Q_{max}(7.75\ MW) = Q_{max}(4\ MW) \cdot \frac{E_{cath}(7.75\ MW)}{E_{cath}(4\ MW)}$$

separately for the Gaussian and the flattop laser temporal profiles. In this formula, $E_{cath}$ corresponds to the accelerating cathode field for each RF gun power setting at the moment of emission, which is at the MMMG for each case. The results of the fit are shown in figure 4, and the fit parameters are summarized in Table 2. From the dimensionless squared halo-core ratio $\eta=1.17$ found from the curve fit, and assuming that the laser radial profile radius of the core is $R_c=0.6\ mm$ (½ of the BSA setting, 1.2 mm), the resultant $\sigma_r$ is 0.46 mm, indicating that the laser beam halo is significantly larger than the inferred from figure 3.

**Table 2.** Fit parameters for the model (2-6) with $\xi = 0.98$, and $\eta = 1.17$ for flattop (17 ps FWHM) and for Gaussian (2.7 ps FWHM). The overall $\chi^2$ of the fit is 53.5, the reduced chi-squared statistic yields $\chi^2_{red} = 0.73$, the number of degrees of freedom $\nu=N_{points}-N_{fit.par} - 1 = 73$.

| Laser temporal profile | RF peak power (MW) | QE (%) | $Q_{max}$ (pC) | $\chi^2 = \Sigma \frac{(meas-fit)^2}{meas.error^2}$ |
|---|---|---|---|---|
| Flattop | 7.75 | 8.36 | 673 | 21.5 |
| Gaussian | | | 445 | 16.7 |
| Flattop | 4.00 | 8.01 | 432 | 5.2 |
| Gaussian | | | 285 | 10.1 |

Based on the results of this fit the ratio of the space charge limiting density (or limiting charge) when comparing profiles for the same gun rf power (cathode accelerating field) is

$$\frac{\sigma_{scl}(flat-top)}{\sigma_{scl}(Gaussian)} \approx 1.51 \quad (7)$$

This ratio shows that the effective space charge limit differs for the considered temporal profiles, but the difference is significantly smaller than the ratio of the laser pulse length (17 ps FWHM / 2.7 ps FWHM ~ 6). The reason may be explained by different longitudinal expansion rates of the bunch due to space charge forces and cathode accelerating field at the moment of emission. The longitudinal expansion is stronger for the Gaussian pulse than for the flattop pulse, while the transverse expansion is very similar for both due to the initial transverse size being almost identical. ASTRA simulations show that the electron bunch rms length ratio when the tail of the bunch has just left the cathode surface for the 7.75 MW case is $\frac{\Delta z_{flattop}}{\Delta z_{Gaussian}} = \frac{\sim 6.0\ mm}{\sim 4.0\ mm} = 1.5$, and for the 4 MW case is $\frac{\Delta z_{flattop}}{\Delta z_{Gaussian}} = \frac{\sim 6.2\ mm}{\sim 4.4\ mm} = 1.4$, which are closer to the value found with the model (see equation 7). Form factors play also a role in the longitudinal expansion of the electron bunches due to the space charge effect, even for bunches having the same rms length, $\sigma_z^{e-bunch} = f\left(\sigma_z^{laser}, \sigma_{xy}, E\right)$. In addition, longitudinal phase space tomography measurements [10] show that the bunch length ratio is ~ 1.6 when the charge from the flattop pulse is 1 nC, and the charge for the Gaussian pulse is 0.7 nC ($Q_{flattop}/Q_{Gaussian}$ ~ 1.4).

Figure 4 shows reasonable agreement of the semi-analytical model for both flattop cases, but quite poor agreement for both Gaussian cases in the transition region between the SC and the QE limited emission regimes. Observing in more detail, the results in figure 4 reveal that the measured charge versus laser energy curve for setup 1 is much stronger saturating than the yield of the modeling, whereas the opposite case – setup 4 – shows the opposite behavior. These discrepancies could be due to the dependence of the space charge density limit on the cathode accelerating field, which is assumed to be $\sigma_{scl} \propto E_0 \sin\varphi$. A more complicated dependence seems more adequate considering the laser temporal profile, implying transient emission and image charge effects strongly dependent on the laser pulse temporal profile. These considerations have being addressed with additional laser transverse distributions characterization and bunch charge measurements compared to ASTRA simulations.

Figure 5 shows a set of transverse laser distributions captured with the UV sensitive CCD camera and processed with the video client software AVINE [11]. The presence of halo (blue) around the core (green) is clear from the pictures, as well as core inhomogeneity characterized by concentric diffraction patterns and, for larger BSA settings, intensity asymmetry caused when the original laser intensity distribution is non-homogeneous. This is due in part to diffraction effects leading to radial modulations and deviation from the designed radial flattop shape. Some optical elements (several lenses and mirrors, and a beam splitter) were placed in the vicinity of the Fourier plane of the BSA-to-photocathode laser beam imaging system. Additionally their apertures (5 cm diameter) truncate high frequency spatial harmonics. These perturbations are most pronounced for smaller spot sizes when the image in the Fourier plane has rather large dimensions.

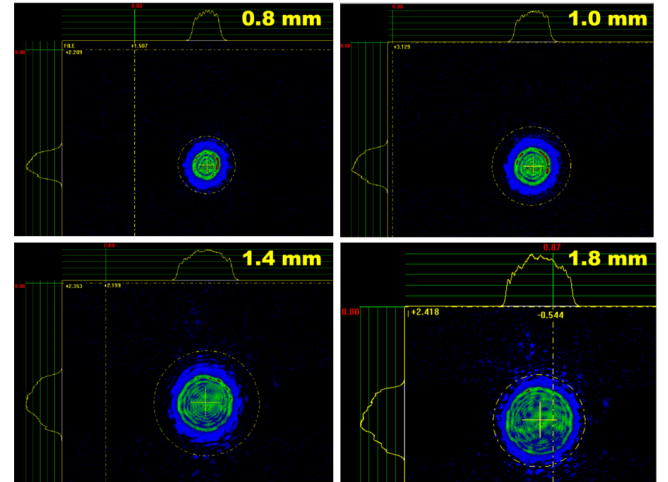

**Figure 5.** Pictures of the laser transverse distribution as captured by the CCD camera software for the indicated laser beam shaping aperture (BSA) settings. Projections from a circular region of interest are shown as well. The images were taken with background "envelope method" subtraction.

A procedure was developed to extract the laser radial profile from images like those shown in figure 5, and also to quantify the amount of halo. By means of a MatLab script, the raw image data and the background data are loaded into two matrices, each with $i \cdot j = 1040 \cdot 1392$ cells corresponding to the CCD pixel array. The value assigned to each $cell_{ij}$ corresponds to the $pixel_{ij}$ intensity registered by the CCD camera and processed by the visualization software. Typically, 20 background frames and 20 image frames make a captured image-background pair. The background subtraction is performed using the "envelope" approach, in which each background $pixel_{ij}$ is assigned the maximum value found in all 20 frames for that particular pixel: $pixel_{ij}^{backgraound} = \max\left(pixel_{ij}^{background}\right)_{L=1..20}$. Then, this value is subtracted from the corresponding image pixel averaged over the 20 frames. If the subtraction $\left\langle pixel_{ij}^{image}\right\rangle - pixel_{ij}^{background}$ renders a negative value, then a value of zero is assigned to the pixel. In addition, the signal from each frame was analyzed separately, and then the results were averaged over the 20 frames. The difference between the frame-by-frame data compared to the averaged image fit is negligible, for the rms beam size the difference is less than 1%, for $R_c$ ~0.04%, and for $\sigma_r$ ~0.4%. A second MatLab script reads the resulting matrix, finds the center of mass in the matrix, makes a horizontal cut across the center and loads the values into a vector. The process is repeated over a series of azimuthal angles (typically, 36 divisions per quadrant for a total of 144 cuts) yielding a rotationally averaged radial profile, which consists of just two data columns, the radial position and the averaged value of the intensity at each radial position. Laser radial profiles obtained in this manner are shown in figure 6. Two features can be observed, a) the laser profile is not uniform flattop (e. g. red dotted profile for the 0.8 mm BSA) as intended to be, but rather shows Gaussian-like decaying halo, and b) the core is far from being a smooth flattop radial profile; the concentric rings diffraction pattern observed in the CCD laser distribution images (Fig. 5) is easily seen as intensity oscillations.

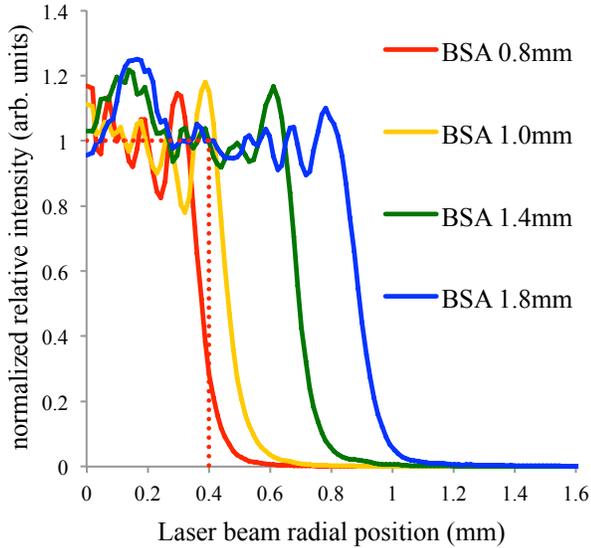

**Figure 6.** Laser radial average profiles extracted from the cathode-imaging plane for various beam shape aperture (BSA) settings. The relative intensity has been normalized with respect to the average in the flattop region for each BSA. The dotted lines represent an ideal BSA 0.8 mm uniform radial profile.

The obtained laser radial profile composed of a flattop core with Gaussian halo can be described by equation (2) as shown in figure 7 for the BSA=0.8 mm.

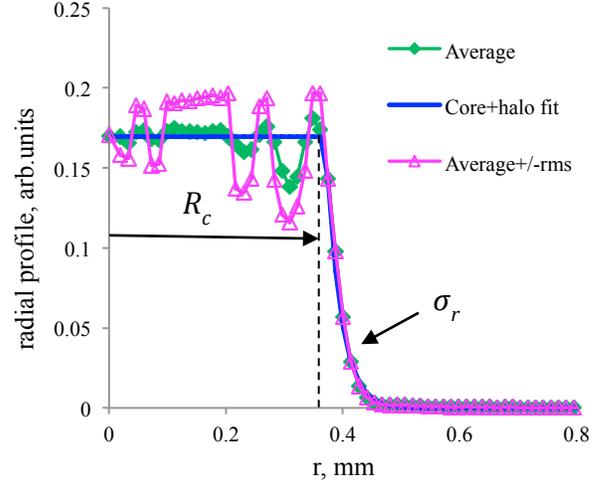

**Figure 7.** Illustration of the fitting parameters $R_c$ and $\sigma_r$ to the laser radial profile data according to equation 2. The core + halo radial profile corresponds to fit parameters $R_c=0.34$ mm, $\sigma_r =0.13$ mm and resulting XY rms= 0.199 mm (identical to the measured value) for BSA=0.8 mm

Once the radial profile is obtained, the area of the core and the area of the halo can be calculated, to a first approximation assuming radially uniform core, by:

$$A_{core} = \pi R_c^2, \qquad (8)$$

$$A_{halo} = 2\pi \int_{R_c}^{\infty} e^{\frac{R_c^2 - r^2}{2\sigma_r^2}} r\, dr = 2\pi \cdot \sigma_r^2 \qquad (9)$$

Figure 8 shows that the halo to core ratio is higher for smaller BSA diameters, confirming what is intuitively observed in the laser distribution images in figure 5.

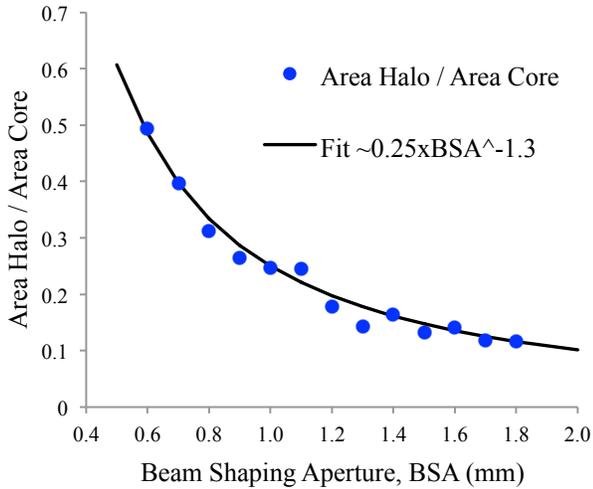

**Figure 8.** The ratio halo to core areas is calculated with equations (8) and (9) for an extended set of experimental BSA values.

*Analysis of the effect of halo on extracted charge*

To study the effect of laser transverse distribution halo on the extracted charge, a series of measurements as a function of laser pulse energy were taken for the BSA set illustrated in figure 9. The logarithmic vertical scale reflects the corresponding dependence in formulas (5) and (6).

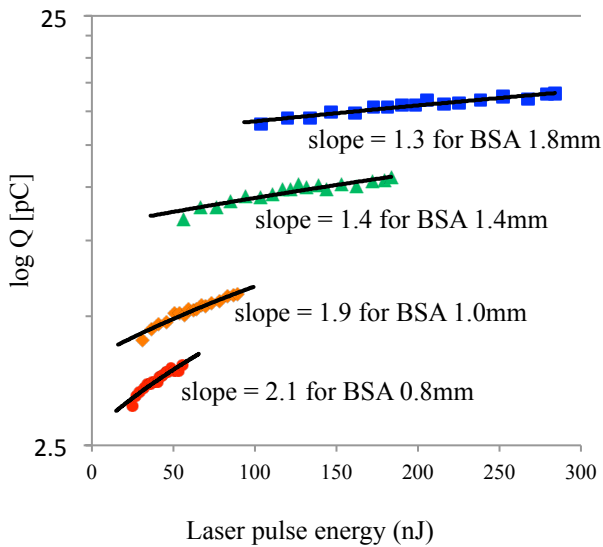

**Figure 9.** Bunch charge as a function of the laser pulse energy for the 3.5 ps FWHM Gaussian temporal laser pulse at the indicated BSA settings with RF gun power = 6 MW and RF phase $\varphi=90°$. The scale of the vertical axis is logarithmic to make more noticeable the slope for the various settings.

The slope in each data set indicates that charge continues to be produced in the SC limited emission regime with increased laser pulse energy, even though the charge from the core has saturated. The slope seems to be stronger for smaller BSA settings, suggesting that more charge is produced in the halo of the laser transverse distribution relative to the core, as shown in figure 10 for two additional RF gun power settings at RF phase $\varphi=90°$.

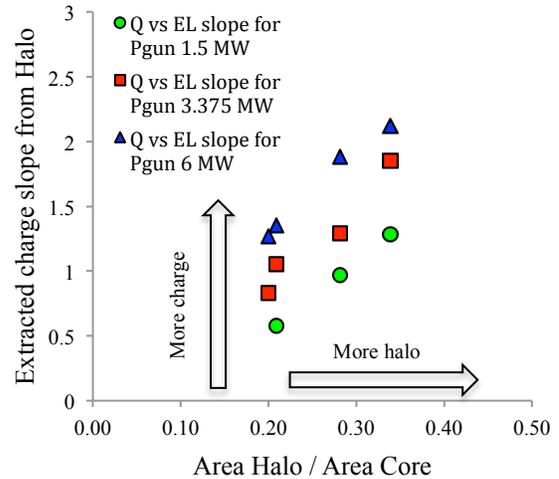

**Figure 10.** Extracted charge slope from laser transverse distribution halo as a function of the ratio Area Halo / Area Core for laser pulse energy for the 3.5 ps FWHM Gaussian temporal laser pulse at the indicated RF gun power settings with RF phase $\varphi=90°$.

This result suggests that:
a) The observed increase in the measured charge in the saturated region is induced by halo in the laser transverse distribution, and
b) Confirms that the amount of halo is larger for smaller BSA diameters due to aforementioned diffraction effects in the laser beam line.

*ASTRA simulations with core + halo particle distributions and comparison with experimental data*

If the presumed homogeneous flattop transverse laser profile is used as input particle distribution in ASTRA with the measured rms size (red profile in Fig. 11), the accelerated charge as a function of input charge (scaled to the laser pulse energy) saturates in the SC limited emission regime (red trace in Fig. 12). If an arbitrarily larger rms laser spot sizes is chosen for the ASTRA input particle distribution (orange distribution in Fig. 11), the saturated charge (orange trace in Fig. 12) does not match the experimental data trend (green circles Fig. 12), it only shifts vertically.

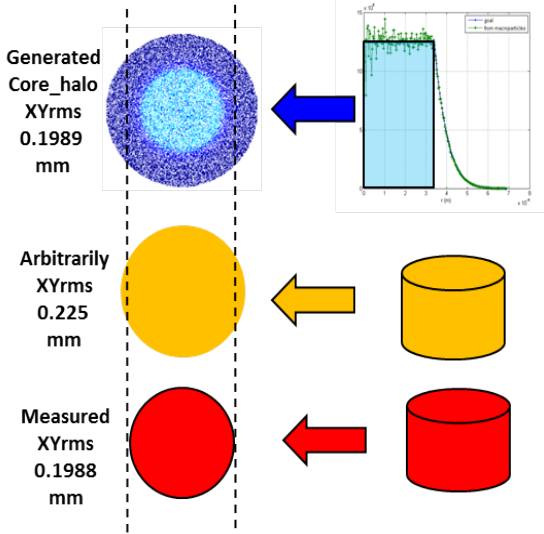

**Figure 11.** Customized radial profile based on fitting parameters to the measured laser radial profile (top rigth), and resulting particle distribution with the macro-particle charge scale accordingly to equation (2) as shown by the post-processing ASTRA software.

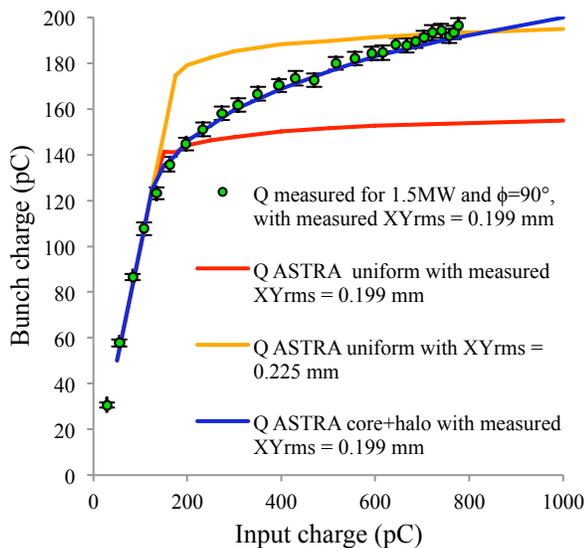

**Figure 12.** Measured (green circles) and simulated (solid lines) charge Q as a function of input charge scaled to laser pulse energy for setup 5 in Table 1. Red trace: ASTRA simulation with homogenous radial profile and XY rms set to measured value of 0.199 mm. Blue trace: ASTRA simulation with core + halo radial profile with fit parameters $R_c=0.34$ mm, $\sigma_r=0.13$ mm and resulting XY rms= 0.199 mm (identical to the measured value). Orange trace: ASTRA simulation with homogenous radial profile XY rms arbitrarily set to a larger value than the measured.

Taking into account the presence of halo based on our characterization of the laser transverse distribution, one can then create customized input particle distributions for the ASTRA simulations composed of a radially homogeneous core with a Gaussian-like decaying halo (see equations 2 and 3). A different MatLab script is utilized to create the new core + halo input particle distribution. The script takes an initially homogeneous radial distribution and scales the macro-particle charge accordingly to $R_c$ and $\sigma_r$ found from the fit parameters in equation (2) to the measured laser radial profile (Fig. 7). When the generated core + halo input distribution is implemented in ASTRA (Fig. 11, top), the simulation results are in close agreement with extracted charge measurements as shown by the blue trace in figure 12, where the simulation result curves have the same set of laser and RF gun parameters for setup 5 (Table 1), with the exception of the shape of the radial profile.

*Sensitivity of the core + halo model implemented in ASTRA to the radial profile fit parameters*

For a given set of RF gun parameters, the behavior of the measured charge vs. input charge curves depends on the radial laser profile parameters $R_c$ and $\sigma_r$ that are utilized to generate the ASTRA core + halo input particle distributions. The fitting to find those parameters is influenced by several factors:

a) Uncertainties in the data capture by the UV CCD camera recording the laser transverse distribution due to the camera sensitivity, the dynamic range, background subtraction, etc. For example, if the laser beam intensity is low, the CDD might be insensitive to photons in the halo that fall below the detection threshold. If the laser intensity is high, the image of the core saturates yielding a larger diameter than the actual spot size on the photocathode, and yielding larger halo than that determined with low laser intensities.

b) Uncertainties in fitting $R_c$ and $\sigma_r$ to the obtained intensity radial profile, in particular as a function of laser intensity. Observations of radial profiles suggest that the amount of halo increases with laser pulse energy, but detailed measurements could not be performed due to camera saturation issues.

c) Uncertainty in the measurements and estimations of the temporal profile of the photocathode laser pulse.

d) Additional uncertainties are introduced by the laser system transport due to diffraction effects resulting in a ring-like structure of the laser transverse distribution at the cathode, plus potential inhomogeneities induced by vacuum viewport and vacuum mirror coupled to inhomogeneities in the cathode QE distribution.

Figures 13 and 14 illustrate the sensitivity of the core + halo model implemented in ASTRA to the radial profile fit parameters when the size of the core $R_c$ is changed by 10% from the value that fits the measured radial profile (blue curves, Figs. 13 and 14), while $\sigma_r$ is adjusted for each $R_c$ value to maintain the rms spot size within 1% of the measured value, 0.199 mm corresponding to BSA=0.8mm. If $R_c$ is reduced, then $\sigma_r$ needs to be increased resulting in more halo, and more halo means stronger increase of the charge in the SC limited emission region (see orange curves in figures 13 and 14). Notice that in this case, the core + halo model fits the experimental data in the transition regime but not in the SC limited emission regime.

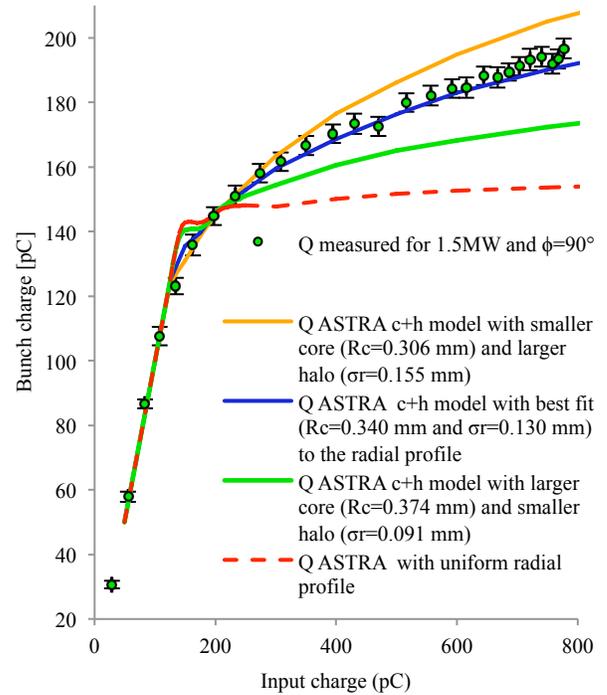

**Figure 13.** ASTRA simulations using core+halo input distributions with indicated $R_c$ and $\sigma_r$ fit parameters adjusted in each case to maintain the rms spot size within 1% of the measured value, 0.199 mm corresponding to BSA=0.8mm in comparison with measured charge for setup 5 in Table 1.

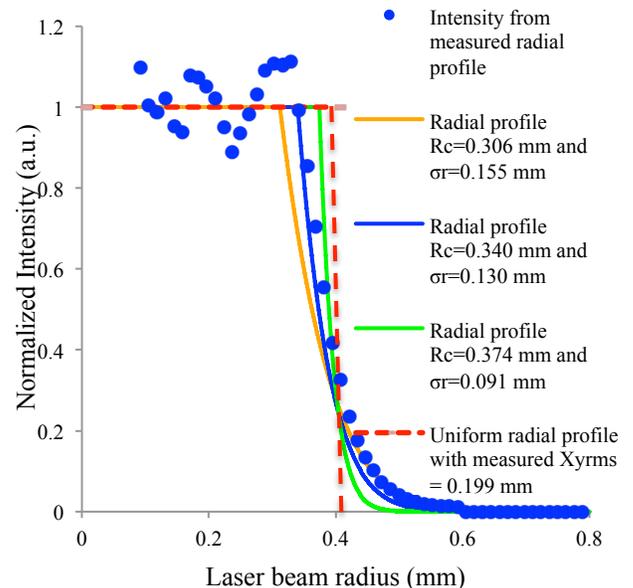

**Figure 14.** Measured laser radial profile data for BSA=0.8 mm in comparison with equation 2 for the $R_c$ and $\sigma_r$ fit parameters listed in figure 13.

In contrast, increasing $R_c$ requires smaller $\sigma_r$ resulting in less halo, therefore less increase of the charge in the SC limited emission regime (see green curves in figures 13 and 14). In this case, the model does not fit the experimental data in the transition regime where the simulation shows the charge saturating before increasing again. The saturation behavior of the green curve in the transition regime is similar to the case with uniform radial profile, as shown by the dotted red curves in figures 13 and 14, indicating that the measured charge saturates in the core before increasing again due to halo as the input charge increases

By utilizing the generated core + halo distribution with $R_c$ and $\sigma_r$ that fit the measured radial profile, ASTRA simulations for setups 5-7 (see Table 1) agree well with the measured bunch charge as shown in figure 15.

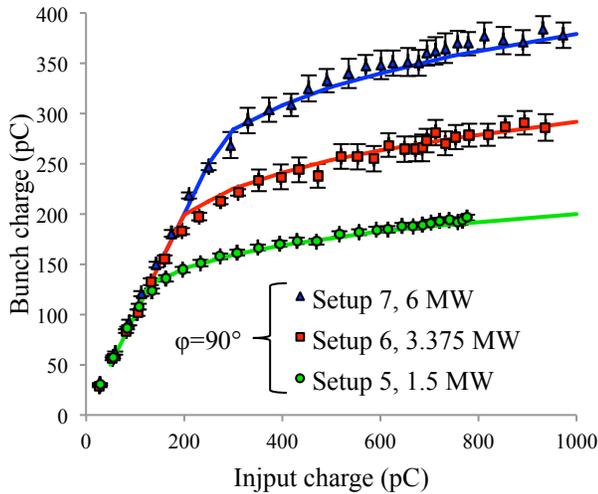

**Figure 15**. Measured charge as a function of input charge scaled to laser pulse energy for setup 5-7 in Table 1. The solid curves show the corresponding ASTRA simulation results using the core+halo distributions with parameters $R_c=0.34$ mm and $\sigma_r=0.13$ mm that fit the laser rms spot size within 1% of its measured value, 0.199 mm corresponding to BSA=0.8 mm.

To illustrate the uncertainty in the injector parameters, a series of experimental runs with 6 MW gun RF power were performed for BSA=0.8 mm at 90, 49 and 30 degrees RF gun phase (setups 8, 9, and 10 respectively in Table 1). The charge measured for each run is compared to the core + halo model implemented in ASTRA for those three phase values indicated by the solid curves in figure 16. The dashed curves show the sensitivity of the model when the phase is changed by +/- 1 degree from the mean value in the core + halo ASTRA simulations.

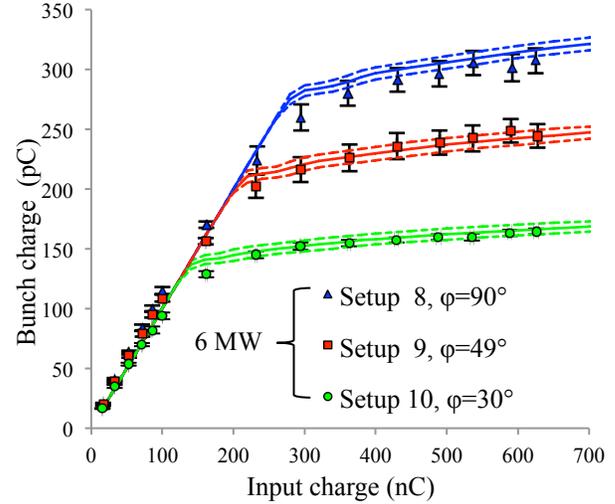

**Figure 16**. Measured charge as a function of input charge scaled to laser pulse energy for setup 8-10 in Table 1. The solid curves correspond to the mean values from the core + halo model implemented in ASTRA with fitting parameters $R_c=0.37$ mm and $\sigma_r=0.10$ mm that fit the laser rms spot size within 1% of its measured value corresponding to BSA=0.8 mm The dashed curves correspond to the model implemented in ASTRA with the gun RF phase at +/- 1 degree from each of the indicated RF phase settings.

The transition region seems to be the regime where results of the core + halo model implemented in ASTRA presented in this work sometimes do not agree well with the experimental data. The discrepancy in the transition regime is not fully understood. For example, the ring structure present in the laser transverse distribution (see figures 5, 6 and 7) was implemented into the core + halo model in ASTRA, but the agreement with the experimental data did not improve, while a less sharp transition was observed if the model was implemented using the average of the transverse distribution intensity +/- the standard deviation around the azimuthal angle (the purple curve in Fig. 7). It should be noticed that the laser pulse used to obtain the curves in figure 16 had the shortest duration, 2 ps as shown in setups 8-10

(Table 1), while the curves in figure 15 were obtained using 3.5 ps long laser pulses (setups 5-7 in Table 1). The only difference in setups 7 and 8 is therefore the duration of the laser pulse. The blue trace in figure 15 (setup 7) shows higher bunch charge for a given input charge when compared to the blue trace in figure 16 (setup 8). This observation indicates higher charge for longer pulses suggesting another signature of the transient character of the emission process. In addition, the azimuthal inhomogeneity implemented in the model (purple curve in Fig. 7) resulted in a smoother curve in the transition between the QE limited and the SC limited emission regimes. Therefore, full 3D simulations are of interest for more precise photoemission process simulations for explaining the remaining discrepancies.

## Conclusions

This work focused on studying the effect of laser transverse halo on extracted bunch charge from $Cs_2Te$ photocathodes in an L-band RF gun as a function of laser pulse energy for a wide range of laser spot transverse sizes, laser temporal pulse profiles, RF gun power and phase settings. Measurements consistently show the bunch charge increasing asymptotically in the space charge limited emission regime, while ASTRA simulations show charge saturation despite direct plug in of experimental laser and RF gun parameters.

In order to understand the source of this behavior, a semi-analytical emission model was applied to charge measurements for ~17 ps-long flattop and for ~2 ps-long Gaussian temporal laser pulses with similar laser radial profiles. The gun RF parameters were set to yield 29 and 45 MV/m accelerating cathode gradient at the moment of emission for each temporal profile. The model is based on a radially homogeneous core with Gaussian-like decaying halo derived from measurements of the laser transverse distribution illuminating the cathode. Although the model agrees reasonably well in the quantum efficiency limited emission regime (low laser pulse energy), the measured bunch charge saturates stronger than the model predictions for the 29 MV/m short Gaussian pulse in the space charge limited emission regime (high laser pulse energy), while the opposite is observed for the 45 MV/m long flattop temporal laser pulses, even though both temporal profiles have very similar transverse distributions. These observations suggest that the asymptotic charge increase in the saturated region is induced by halo present in the laser transverse distribution despite charge saturation in the core of the distribution, and that transient emission and image charge effects strongly dependent on the cathode laser pulse temporal profile.

Characterization of the laser transverse distributions as a function of spot size indicates that the amount of halo-to-core ratio increases as the laser spot size is reduced. The ratio has an inverse power law behavior with the laser spot size. We attribute this behavior to diffraction effects resulting from the existing configuration of the laser beam transport system. Correlating the amount of measured charge in the space charge saturation regime with the halo-to-core ratio suggests a linear relationship between the two.

To test out hypothesis, custom particle input distributions composed of a flattop core with radius $R_c$ and Gaussian-like decaying halo with $\sigma_r$ were generated after fitting these parameters to radial profiles derived from the characterization of the laser transverse distributions. When the core + halo customized input particle distributions are utilized, ASTRA simulations reproduce well the behavior of the measured charge vs. laser pulse energy in the space charge limited emission regime for a wide range of laser spot sizes and RF gun parameters, in contrast to saturation of the accelerated bunch charge when only the presumed homogeneous flattop radial laser profile was used as input particle distribution. However, the core + halo model implemented in ASTRA sometimes overestimates or underestimates the charge measurements.

The systematic limitations of this approach depend on the cumulative uncertainties related to the actual charge transverse distribution on the cathode, which in turn depends on the generated laser transverse distribution, the laser optical transport system to the cathode and on the cathode QE uniformity, on the measurement and characterization of the laser radial and temporal profiles (in particular for higher laser pulse energies where the CCD signal saturates), on the algorithm to derive an average laser radial profile, and on the manual fit of the core + halo parameters from the obtained radial profiles.

Despite these limitations, our analysis confirms that the presence of halo in the laser transverse distribution contributes to production of excess charge as the laser pulse energy increases in the space charge limited emission regime where the charge from the core has saturated.

Although the approach presented in this work attributes the observed charge behavior vs. laser pulse energy to the presence of halo in the cathode laser transverse distribution illuminating the photocathode, improvements of the laser transport system should minimize these effects, therefore rendering a distribution closer to the ideal homogenous flattop radial profile.